\documentclass[authoryear,1p]{elsarticle}
\usepackage{adjustbox}
\usepackage{tikz}
\usetikzlibrary{decorations.pathmorphing,patterns}
\usepackage{graphicx}
\usepackage{dcolumn}
\usepackage{bm}
\usepackage{hyperref}
\usepackage{amsmath}
\usepackage[capitalise]{cleveref}
\usepackage{subfig}
\def\appendixname{}
\usepackage{dblfloatfix}
\usepackage{framed} 

\journal{Advanced Powder Technology}
\begin{document}

\begin{frontmatter}

\title{Triboelectric Charging Model for Particles with Rough Surfaces}

\author[inst1,inst2]{Simon Jantač\corref{cor1}}
\ead{simon.jantac@ptb.de}
\author[inst2,inst4]{Jarmila Pelcová}
\author[inst2,inst4]{Jana Sklenářová}
\author[inst2]{Marek Drápela}
\author[inst1,inst3]{Holger Grosshans}%
\author[inst2]{Juraj Kosek}

\affiliation[inst1]{organization={Physikalisch-Technische Bundesanstalt (PTB)},
            addressline={Bundesallee 100}, 
            city={Braunschweig},
            postcode={D-38116}, 
            country={Germany}
            }
            
\affiliation[inst2]{organization={Univeristy of Chemistry and Technology Prague, Department of Chemical Engineering},
            addressline={Techncká 5}, 
            city={Prague},
            postcode={160 000}, 
            country={Czechia}
            }
            
\affiliation[inst4]{organization={DPI, P.O. Box 902, 5600 AX Eindhoven, the Netherlands}}
            
\affiliation[inst3]{organization={Otto von Guericke University of Magdeburg, Institute of Apparatus and Environmental Technology},
            addressline={Universitätsplatz 2}, 
            city={Magdeburg},
            postcode={D-39106}, 
            country={Germany}}
         	
\cortext[cor1]{Corresponding author}

\begin{abstract}
The triboelectric charging of particles depends on the contact area of the particle and the contacting surface.
Even though the surface topology determines the real contact area, particle charging models do not account for surface roughness.
In this paper, we combine contact mechanics and triboelectrification models to predict the charging of rough particles.
First, a laser confocal microscope measured the statistical descriptors of polyethylene (PE) particles surface topology.
Then, we descriptors particle surfaces by distributing spheroidal asperities on the smooth particle core until the surface roughness reaches the measured value.
The Hertz contact mechanics model predicts the deformation of the asperity-covered particle and the resulting real contact area in dependence on impact velocity.
Finally, we introduced the real contact area into the condenser model for triboelectric particle charging.
The accuracy of the new model predictions was demonstrated by comparing it to a more complex surface reconstructions that account for the fractal surface topology.
Furthermore, the model's predicted particle saturation charges agree well with our shaker experiments and with experimental data in the literature on the charging of plane surfaces.
The developed triboelectric charging model for particles with rough surfaces is simple and requires only standard descriptors of the surface topology;
thus, it suits large-scale simulations of electrifying powder flows.
\end{abstract}

\begin{graphicalabstract}
\includegraphics[width=\textwidth]{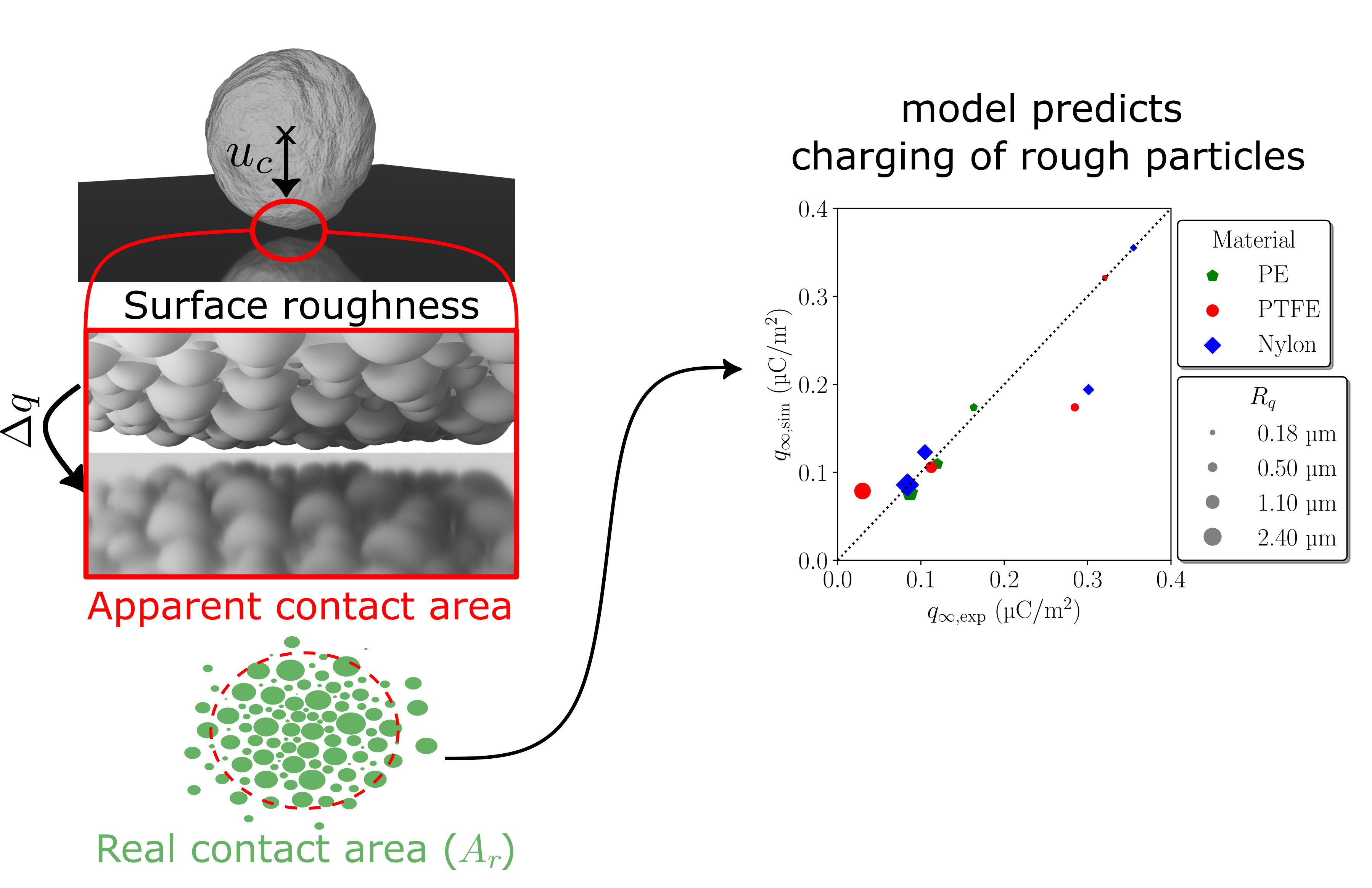}
\end{graphicalabstract}

\begin{highlights}
\item New particle charging model predicts the influence of surface roughness.
\item Requires standard descriptors of surface topology.
\item Agrees well with charging experiments.
\item Suits large-scale simulations of electrifying powder flows.
\vfill
\end{highlights}


\end{frontmatter}

\begin{table*}[!t]   
	
	\begin{framed}
		
		\begin{tabular}{l p{4.5cm} l p{4.5cm}}
			\multicolumn{2}{l}{\textbf{Nomenclature}}
			 \\
			 & \\
			\multicolumn{2}{l}{\textbf{Variables}} \\	
			$\beta$ & Condenser model fitting parameter (-) & $E^*$& Effective Young’s modulus ($Pa$) \\									
			$\Delta q$ & Charge transfer ($C$)	& $F$ & Force ($N$) \\ 													
			$\delta$ & Compression ($m$) & $H$ & Hurst exponent (-) \\														
			$\kappa$ & Dimensionless constant of contact normal pressure of rough surface (-)& $h^{'}_{\mathrm{rms}}$ & Dimensionless root mean square slope of surface (-) \\ 	
			$\nu$& Poisson’s ratio (-)& $k$ & Slope of the power spectrum ($m^5$) \\ 															
			$\sigma$& Stress ($Pa$)&$p_\mathrm{rough}$ & Contact normal pressure of rough surface ($Pa$) \\	 															
			$\varepsilon$& Electric permittivity ($F/m$)& $q_\infty$ & Saturation charge ($C$) \\ 										
			$\xi$& Magnification (-) & $r, \theta, \phi$ & Spherical coordinates ($m, ^{\circ}, ^{\circ}$) \\															
			$A$& Contact area ($m^2$)& $R_a$ & Average surface roughness ($m$)  \\ 															
			$A_c$& Contact area of macroscopic asperity ($m^2$)& $R_q$ & Root mean square of the surface roughness ($m$)  \\ 								
			$A_r$& Real contact area  ($m^2$)	& $S(w)$ & Surface power spectrum ($m^4$)  \\												
			$A_\mathrm{max}$& Maximum contact area & $u_c$ & Normal impact velocity ($m/s$)  \\												
			$A_{r,\mathrm{tot}}$& Real contact area of complete particle surface ($m^2$)& $V$ & Contact potential difference ($V$)  \\		
			$C$& Contact capacity ($F$) & $w$ & Wave number ($1/m$)	\\ 													
			$d$& Separation distance ($m$)	& $x,y,z$ & Cartesian coordinates ($m, m, m$) \\													
			$E$& Young’s modulus ($Pa$) & $z_c$ & Contact gap ($m$)  \\
			\\
			\multicolumn{2}{l}{\textbf{Subscripts}} \\
			$i$ & Variable connected to the \(i\)-th asperity & $p$ & Variable connected to particle \\
			$n$ & Variable's normal component \\
		\end{tabular}		
		
	\end{framed}
	
\end{table*}

\section{Introduction}

Surface asperities reduce the real contact area of two surfaces touching each other.
Therefore, surface roughness significantly influences triboelectrification, which requires physical contact to transfer charge between two surfaces.
However, the available models for particle charging generally neglect surface roughness~\citep{Grosjean2023,MATSUSAKA20105781,Konopka2017}, despite the scientific progress in modeling particle contact mechanics \citep{Persson2002,PERSSON2006201}.
Consequently, simulations of the contact electrification of particles and particle-laden flows generally cannot predict the effect of surface roughness on powder charging~\citep{CHOWDHURY2021104,GROSSHANS2023140918,Sip18,Zhang23}.

Triboelectrification or contact electrification occurs when electric charge spontaneously separates upon contact.
Often, this charge separation is unwanted and impairs industrial processes.
Thus, a numerical model for particles that reflects the effect of surface roughness on triboelectrification would improve the accuracy of large-scale DEM (Discrete Element Method) and CFD (Computational Fluid Dynamics) simulations and help to fight process losses.

Examples of particle-laden flow systems that can electrify and could benefit from simulations considering surface roughness include fluidized bed reactors.
During the synthesis of polyolefins, the accumulation of charge due to triboelectrification causes the formation of wall sheets that reduce heat transfer and degrade the product.
If the powder charges bipolarly, positive and negative particles attracting each other can agglomerate, and the powder lumps~\citep{HENDRICKSON20061041,SOWINSKI20102771,JANTAC2019148,KONOPKA2020713}. 
Further, pneumatic conveying, due to the high conveying velocities, leads to rapid powder charging~\citep{KLINZING201878}.
The sudden release of accumulated charge can cause the ignition of the conveyed powder and dust explosions, especially in downstream units \citep{OHSAWA2017171,Gro24b,Glor03}.
In the pharmaceutical industry, most powders are non-conductive, hence prone to charge accumulation.
Besides unwanted sheeting and the overall complex handling of charged powders, the pharmaceutical industry has to pay attention to bioavailability affected by dissolution rates, often altered by charged agglomerates of active pharmaceutical ingredients and excipients.
Also, the consistency of drug dosing is affected by the segregation of charged powder blends \citep{Alfano2023,Middleton2023,KAIALY2016262}.

\begin{figure*}[tb]
\subfloat[]{%
\label{fig:Topoa}%
\includegraphics[width=0.32\textwidth]{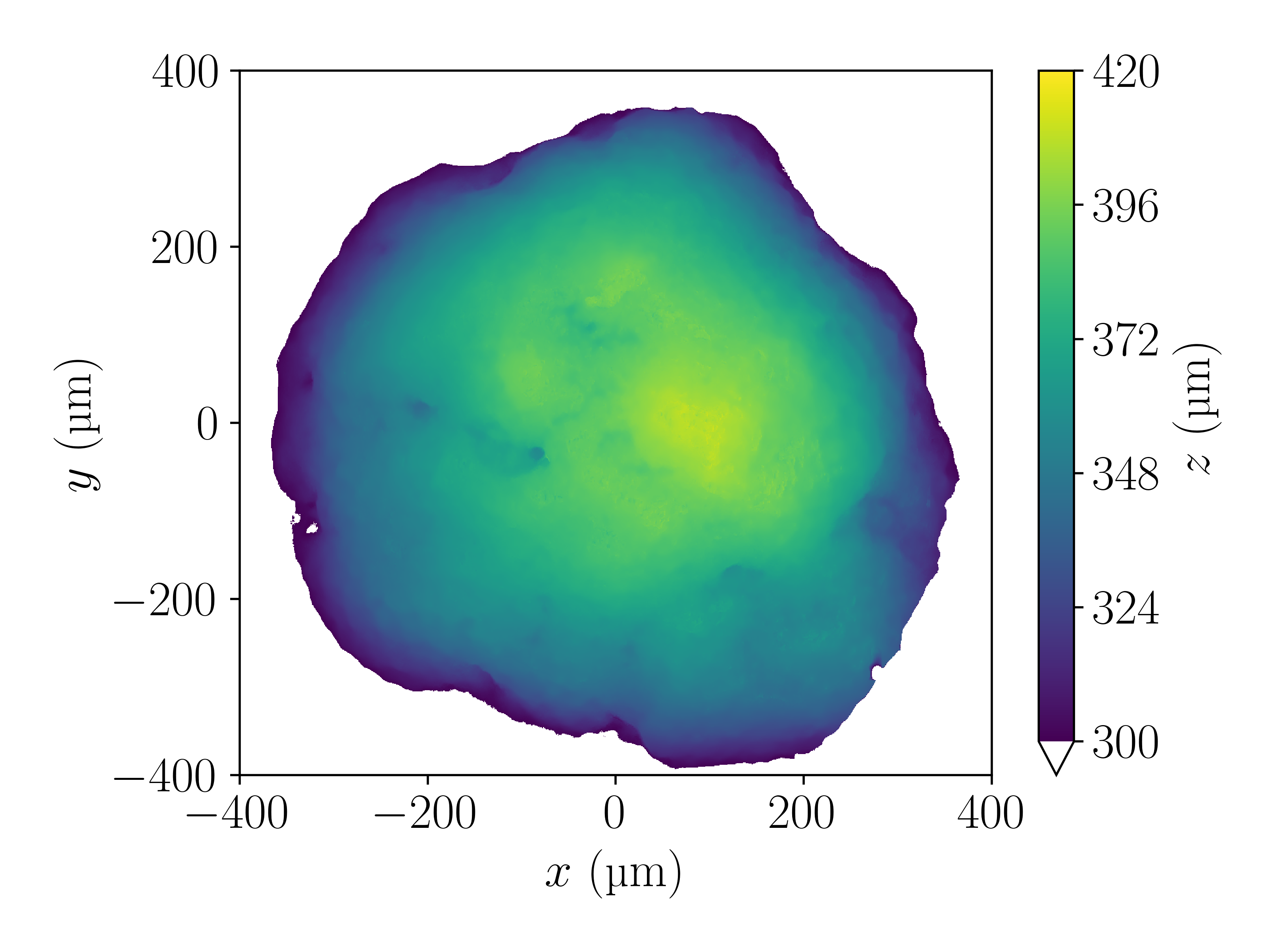}
}~
\subfloat[]{
\label{fig:Topob}%
\includegraphics[width=0.32\textwidth]{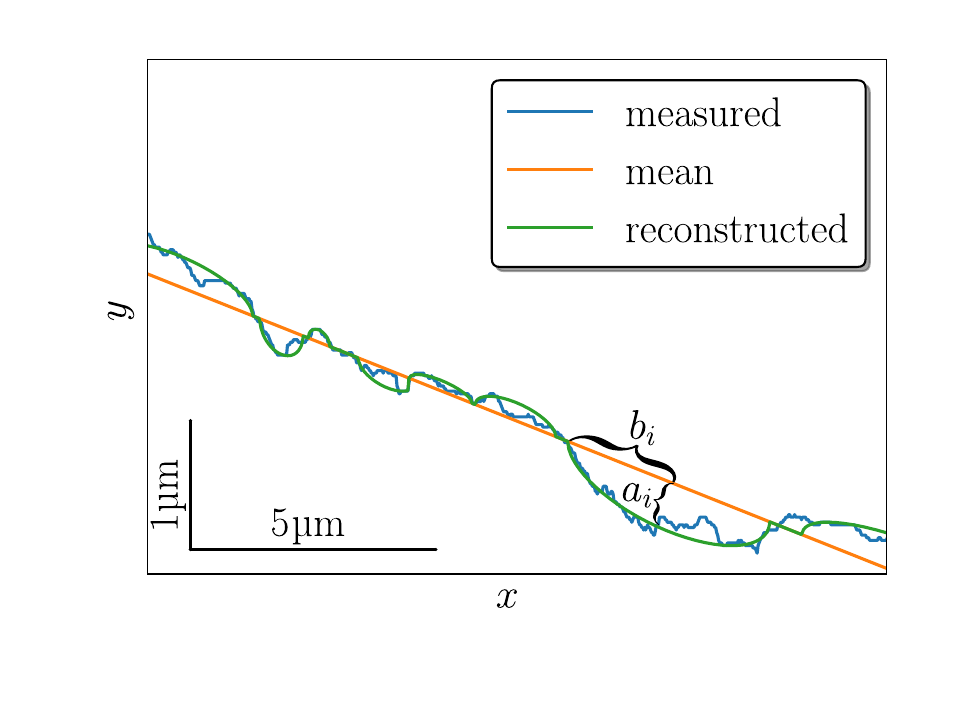}
}~
\subfloat[]{
\label{fig:Topoc}%
\includegraphics[width=0.32\textwidth]{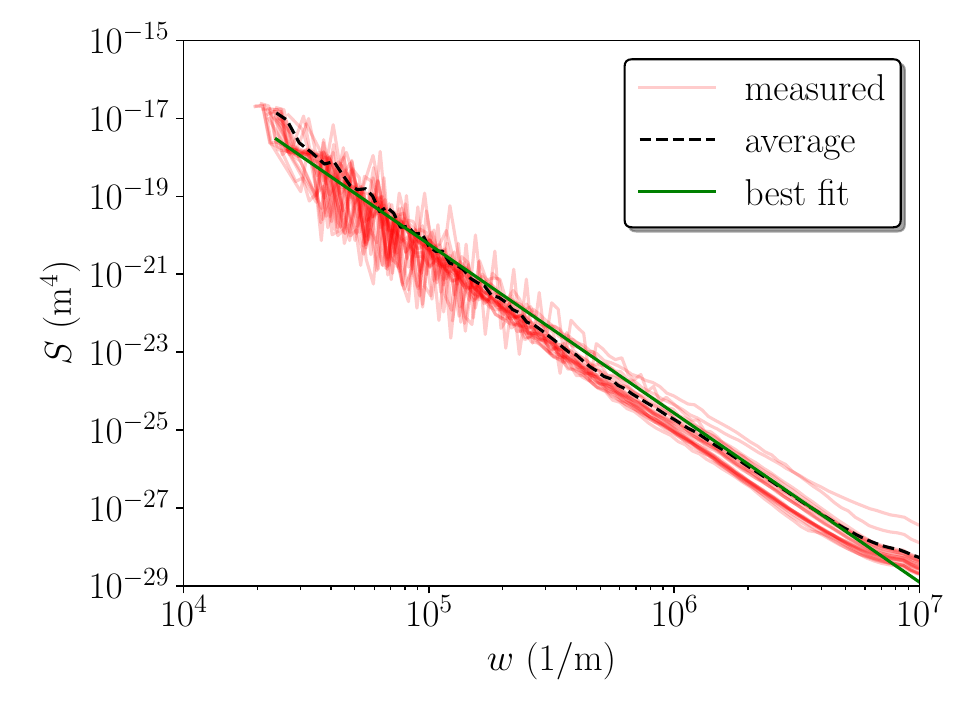}
}
\caption{
(a) Topology of nascent PE particle projected on the $x$-$y$ plane, acquired by a laser confocal microscope;
(b) 2D sample slice of the 3D topology and surface roughness analysis;
(c) Surface roughness power spectrum.
}
\label{fig:Topo}
\end{figure*}

Most triboelectric charging models assume that the charge transferred between a particle and another surface scales with their contact area.
For example, the condenser model~\citep{MATSUSAKA20105781} quantifies the charge transfer $\Delta q$ by
\begin{equation}
\label{eq:condenser}
\Delta q = \beta V C \quad \text{with} \quad C=\varepsilon A/ z_c \, ,
\end{equation}
where $V$ is the contact potential difference, which is the driving force of the charge transfer, and $\beta$ is a fitting parameter.
Furthermore, $C$ is the capacity, which depends on the electric permittivity, $\varepsilon$, of the gas filling the contact gap $z_c$, and the contact area $A$.
Thus, except $A$, all variables on the right-hand side of \cref{eq:condenser} are constants for given surface materials, and the total charge transfer scales linearly with $A$.

Even though the condenser model is valid for the contact between conductors, it has been applied also to insulating surfaces as well~\citep{Wat03,Kol89}.
The product $\beta V$ is usually fitted to experimental data because it cannot capture all the parameters influencing charging.

Patch models address the charge exchange between insulators.
Therein, the charge transfer still occurs at the contact area, but the contact area is discretized into smaller sites of fluctuating properties.
The fluctuations can be caused by defects in the crystalline structure, impurities, or adsorbed species.
Patch models can explain the polarity reversal due to the asymmetric friction \citep{Grosjean2023a}, but the computational domain size is limited to a few micrometers.
Averaging the local patches allowed to increase the possible domain size while preserving some properties of the patch model~\citep{Lacks2008,Konopka2017,Zhang2024}, thus enabling its use in large-scale simulations \citep{Jantac2024}.

Despite the differences between condenser and patch models in terms of their application, underlying assumptions and predictions, both model classes agree that the transferred charge scales with the contact area.
The model assumption $q \propto A$ is backed up by experiments~\citep{IRELAND2010189,IRELAND2010199,IRELAND2012524}.
For example, the impact charge of glass particles scales linearly with the impact velocity~\citep{LESPRIT2021103605}, which in turn scales with the contact area.
Also, the impact charge of smooth, soft particles, which completely flatten during impact, is proportional to the contact area~\citep{Shuji_Matsusaka_2000}.
Overall, for most materials, the impact charge is most likely a linear function of the impact velocity~\citep{WATANABE2007149}.
However, in the cited experiments, the particles were smooth, their surface roughness was not measured, or their impact velocity was so big that the asperities on their surfaces were probably wholly deformed.

Simulations of powder charging typically compute the contact area using a Hertzian model, which assumes spherical particles with perfectly smooth surfaces.
However, for rough particles, the real contact area is only a fraction of the apparent contact area \citep{Johnson_1985}.
A few experimental studies showed that triboelectrification depends not on the apparent but the real contact area~\citep{HU201985,Ge2023b,Ire08}.
The real contact area is especially important for insulators, where the transferred charge does not spread beyond the area of immediate contact;
thus, surface roughness can limit the charge transfer to a fraction of what would be transferred by smooth surfaces~\citep{Masui_1984}.
However, as stated above, particle charging models have not yet considered the reduction of the contact area during collisions due to surface roughness.

In this paper, we connect contact mechanics and particle triboelectrification models to propose a model that predicts particles' charging depending on their surface roughness.
\Cref{sec:model} details the model development, \cref{sec:sim} reports on simulations using the model and comparison to experiments, and \cref{sec:con} concludes the paper.

\section{Triboelectric model for rough particles}
\label{sec:model}

This section documents the development steps and parts of the triboelectric model.
First, the surface topology of real particles is measured, and the required statistical descriptors are evaluated (\cref{sec:Toposec}).
These descriptors are used to model particles with statistically identical surface roughness (\cref{sec:stat}).
Then, we compute the contact mechanics of the modeled surfaces and the real contact area (\cref{sec:CA}).
Finally, we implement the real contact area in a charge transfer model (\cref{sec:CT}).

\subsection{Surface topology of real particles} 
\label{sec:Toposec}

To obtain input data for the numerical model regarding the surface roughness of real particles, we analyzed the topology of twenty nascent polyethylene (PE) particles of diameters between 0.8~mm and 1~mm, see \cref{fig:Topo}.
While PE particles produced on the industrial scale have a wide range of shapes and random surface topologies~\citep{McKenna2010,Alizadeh2018}, we found mostly spherical asperities.
\Cref{fig:Topoa} plots the topology of one particle obtained by a laser confocal microscope (Keyence VK-X3000, 10$\times$ magnification, pixel size of 691~nm).
Similar 2D topological data can be obtained by profilometers or by slicing 3D topological data, e.g., from atomic force microscopes (AFM), into 2D profiles.
\Cref{fig:Topoa} visualizes approximately the top half of the particle, where the measured $z$ coordinate of the surface is above 300~µm, projected on the $x$-$y$ plane.
Slices through this topological data yield one-dimensional profiles of the surface, one example being plotted in \cref{fig:Topob}. 

From these profiles, standard roughness metrics, like average roughness $R_a$, and root mean square roughness $R_q$, can be obtained.
Further, we receive the surface roughness power spectrum $S(w)$, where $w$ is the wave vector of the rough surface \citep{PowerSpec}.
\Cref{fig:Topoc} depicts the power spectrum of all analyzed particles and their mean.
Fitting the equation
\begin{equation}
\label{eq:power}
S(w)=k w^{-2(H+1)}.
\end{equation}
to the measured spectrum yields the slope $k$, and the Hurst exponent $H$, of the averaged power spectrum \citep{PowerSpec}.
The surface roughness power spectrum is required by the contact model of \citet{PERSSON2006201}, which is employed in \cref{sec:CA}.
For the particle surfaces depicted in \cref{fig:Topo}, the topological analysis yielded $R_a = 1.8 \ $\textmu m, $R_q =2.1 \ $\textmu m, $k=28.9$, and $H=1.13$. 

Further, the topological data delivers the statistical distribution of macroscopic asperities.
The procedure for finding the distribution is as follows:
The macroscopic asperities are approximated by spheroids deposited on the smooth particle core.
First, we compute the mean of the surface profile.
Then, we add a layer of macroscopic asperities to the mean.
The width of each asperity, $b_i$, is obtained from the intersection of the measured profile and the mean surface;
their height, $a_i$, is estimated from the highest or lowest position between those intersections.
The center of the macroscopic asperity is placed at the midpoint between the intersections. 
This procedure's output is the distribution of the shapes and sizes and the number density of the macroscopic asperities.

As a final note, if only standard metrics of surface roughness ($R_a$ or $R_q$) are available but no measurements of the surface topology, one can assume that the asperity sizes are Gaussian distributed.
We will test this assumption below.

\subsection{Surface reconstruction of the model particle}
\label{sec:stat}

In the second step, we set up a model of a particle with statistically equivalent surface roughness as the real particle analyzed in \cref{sec:Toposec}.
Thus, the surface reconstruction uses the data obtained from the topological analysis.
The model particle aims to have the same asperity number density, size and shape distribution as the real particle and the same $R_q$ and $R_a$, the latter with a maximal deviation of 15\%. 

The surface reconstruction starts with a smooth core particle.
Then, macroscopic asperities are deposited on the core's surface.
The core particle's radius ($r_p$) is one mean asperity size below the mean surface depicted in \cref{fig:Topob}.
Thus, the seeded asperities have the height $2 a_i$ to reach the same mean surface as the real particle.

\begin{figure}[tb]
\centering
\subfloat[]{
\label{fig:asperity_coor}
\includegraphics[width=0.27\textwidth]{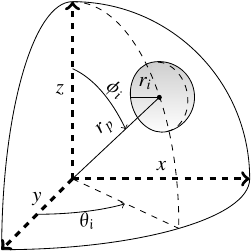}
}
\subfloat[]{
\label{fig:contact_mechanics}
\includegraphics[width=0.27\textwidth]{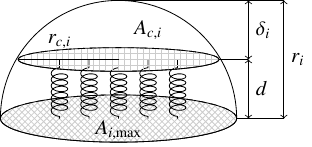}
}
\caption{
(a) Spherical ($\phi_i$, $\theta_i$, $r_p$) and Cartesian ($x_i$, $y_i$, $z_i$) coordinate system of the particle.
For clarity, one-eighth section of the core particle and one deposited macroscopic asperity (grey) is depicted.
(b) Coordinate system of a macroscopic asperity that is compressed by $\delta_i=r_i-d$.
The springs illustrate a linear contact mechanics model, whereas the study tested different models.
}
\end{figure}

The macroscopic asperities can be half-spheroids or half-spheres (see \cref{sec_A1}), where half-spheres simplify the contact mechanics model in \cref{sec:CA}.
The coordinate system of the particle and one deposited spherical asperity is given in \cref{fig:asperity_coor}.
Each asperity has a position ($\phi_i$, $\theta_i$, and $r_p$ in the spherical or $x_i$, $y_i$, and $z_i$ in the Cartesian coordinate system), radius ($r_i$), and elastic properties (Young's modulus, $E_i$, and Poisson's ratio, $\nu_i$).

\begin{figure*}[tb]
\centering
\subfloat[]{\includegraphics[width=0.27\textwidth]{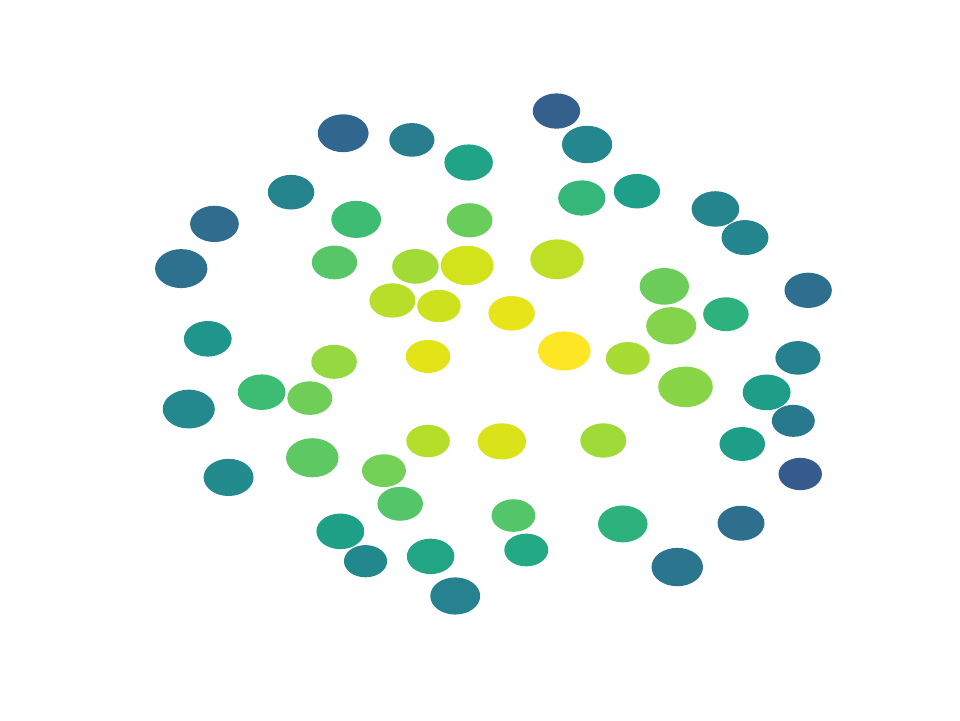}}
\hspace{-4mm}
\subfloat[]{\includegraphics[width=0.27\textwidth]{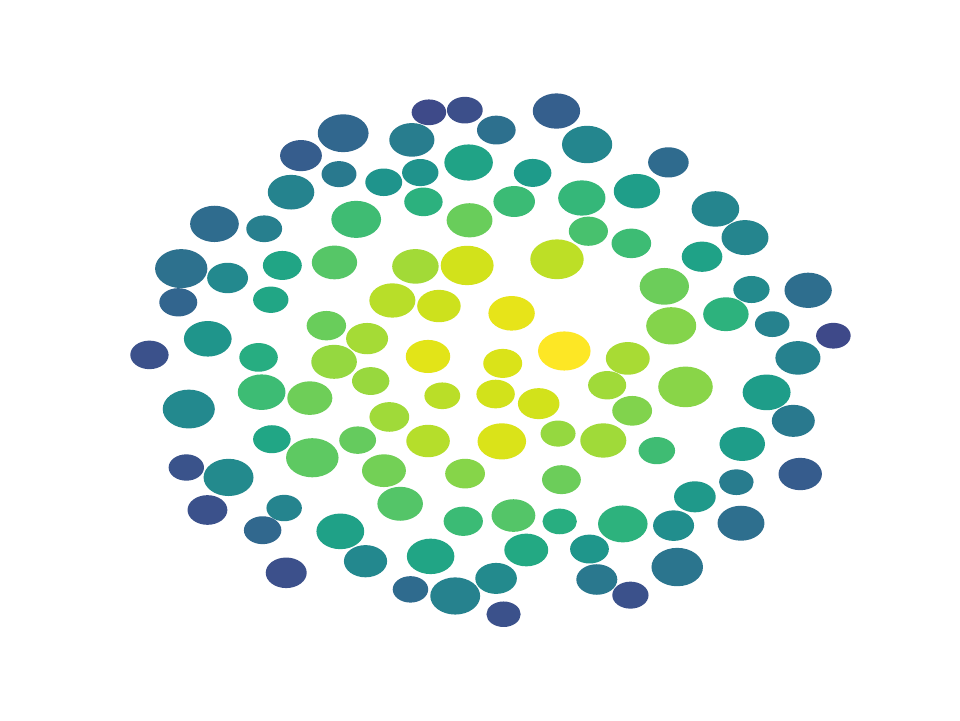}}
\hspace{-4mm}
\subfloat[]{\includegraphics[width=0.27\textwidth]{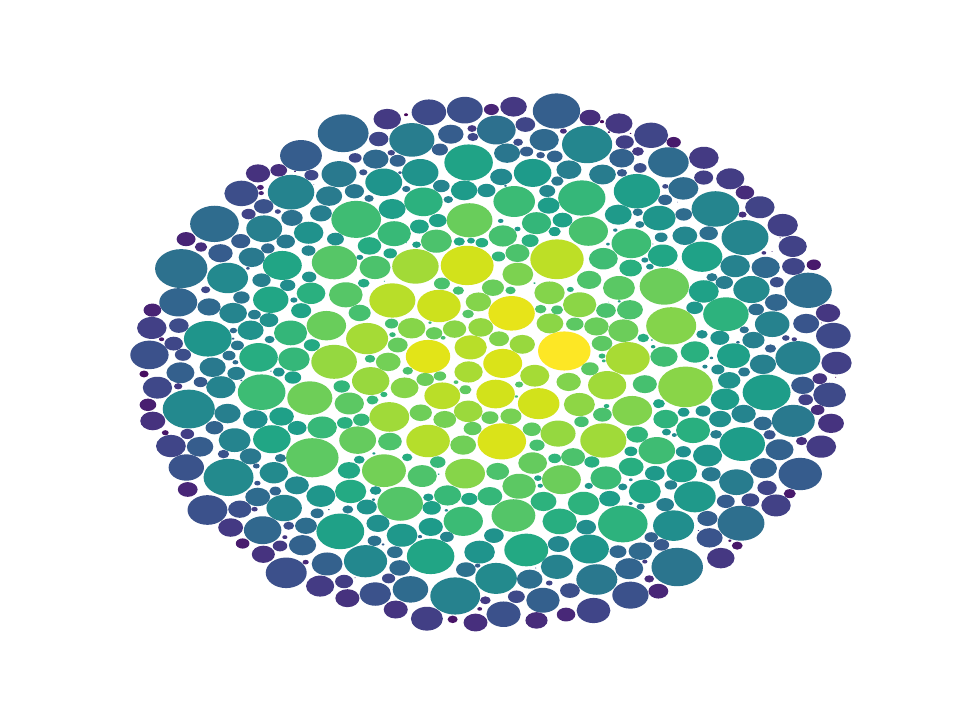}}
\qquad
\adjustbox{trim=0cm 0cm 0cm 2.55cm}{%
\includegraphics[width=0.06\textwidth]{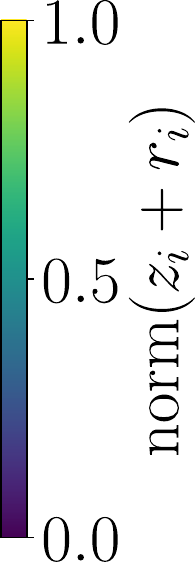}%
}
\caption{
Surface reconstruction of the modeled particle's contact area viewed from the surface normal.
The color marks the normalized position of the asperity apex (norm($z_i+r_i$));
0 marks the lowest and 1 the highest asperities.
The snapshots depict the sequence of the reconstruction procedure:
(a) 1/10 of the largest asperities are seeded;
(b) 1/5 of the largest asperities are seeded;
(c) all asperities are seeded.
}
\label{fig:Reconstruct}
\end{figure*}

In the reconstruction, we randomly place asperities on the particle until we reach the desired $R_a$ and $R_q$.
The resulting asperity number density is slightly lower than in the real particle, because at this point, the modeled surface has only macroscopic asperities.
The effect of the smaller asperities is modeled below based on the power spectrum (\cref{fig:Topoc}).
A further challenge to reaching the required asperity number density lies in the asperities' high coverage of the particle's surface.
To avoid overlapping while reaching a high asperity number density, we first sample the asperity sizes from their distribution until the required coverage ratio is achieved.
Then, a physical location is assigned to the largest asperities (see \cref{fig:Reconstruct}).
Afterward, the smaller asperities fill the voids between the large asperities.
This approach keeps the asperity distribution randomized without bunching near the spherical cap's poles.
Further, seeding the particle only at the contact area tremendously reduces the required memory and computational effort.

\subsection{Contact mechanics model and real contact area}
\label{sec:CA}

After reconstructing the real particle surface with macroscopic asperities, the next step is to develop the contact mechanics of the modeled particle.
We assume that the macroscopic asperities are mechanically independent of each other and have spherical shapes.
The contact scenario is a rough particle colliding head-on with a flat, rigid half-space.
For this scenario, the contact mechanics model gives each asperity's contact area and force.

As depicted in \cref{fig:contact_mechanics}, during the collision, each macroscopic asperity $i$ that is higher than the separation distance $d$ compresses by $\delta_i=r_i-d$.
Then, the contact area of each macroscopic asperity becomes
\begin{equation}
A_{c,i}=\pi r_{c,i}^2 \, ,
\label{eq:MactroContactArea}
\end{equation}
where $r_{c,i}$ is the radius of the contact area.
The total real contact area is the sum over all $N$ asperities in contact, i.e.,
\begin{equation}
A_r=\sum_{i=1}^N A_{c,i} \,.
\label{eq:ar}
\end{equation}

Assuming that the contact force is shared between the asperities and the core particle, the well-known contact mechanics model by Hertz gives the asperity contact force, $F_i$,
\begin{equation} \label{eq:Hertz_asp}
F_i= \dfrac{2}{3} E^* \, r_p^{1/2} \, \delta_{i,p}^{3/2} \, ,
\end{equation}
where $E^*=E/(1-\nu^2)$ is the effective Young modulus and $\delta_{i,p}$ the penetration of the deformed asperity into the core particle.
The contact force taken by the particle is 
\begin{equation}
 F_p=\dfrac{2}{3} E^* \, r_p^{1/2} \, \delta_p^{3/2} \, .
\end{equation}
Finally, the total contact force is the sum of the contact forces of the asperities and the the core particle,
\begin{equation}
F_n=F_p+\sum_{i=1}^N F_i.
\label{eq:SumForce}
\end{equation}

\begin{figure*}
\centering
\hspace*{-2cm} 
\subfloat[]{%
\includegraphics[scale=0.33]{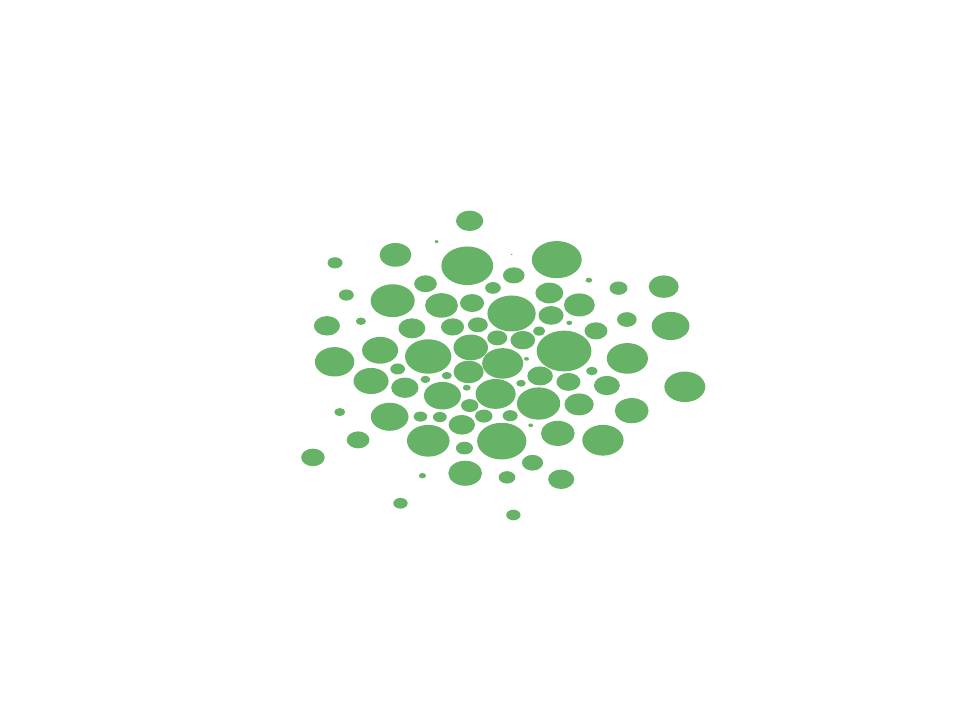}
\includegraphics[scale=0.33]{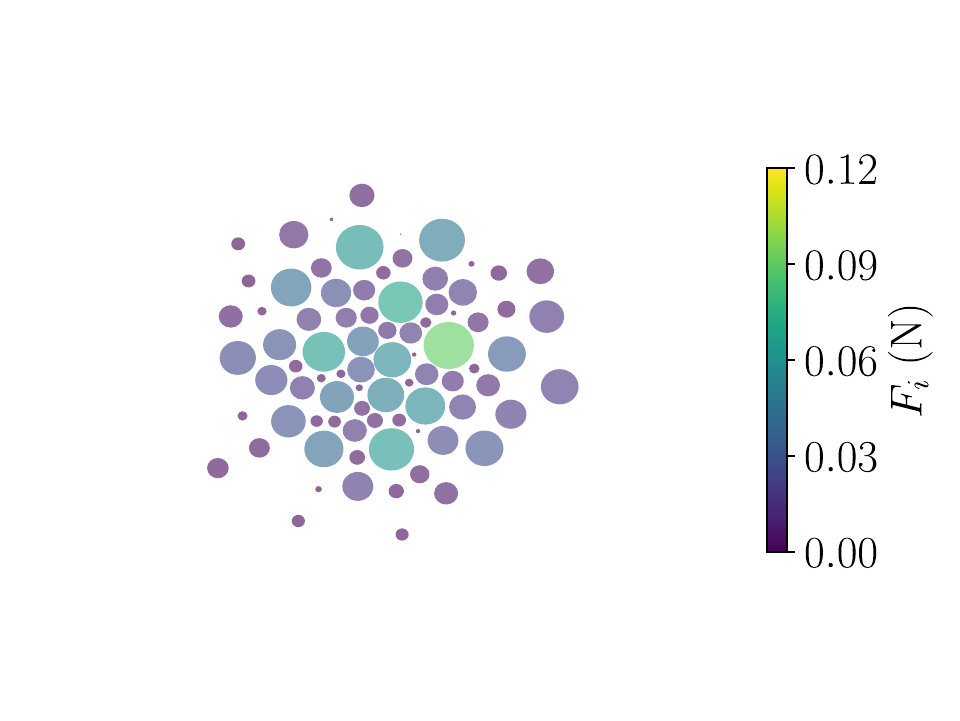}
\includegraphics[scale=0.33]{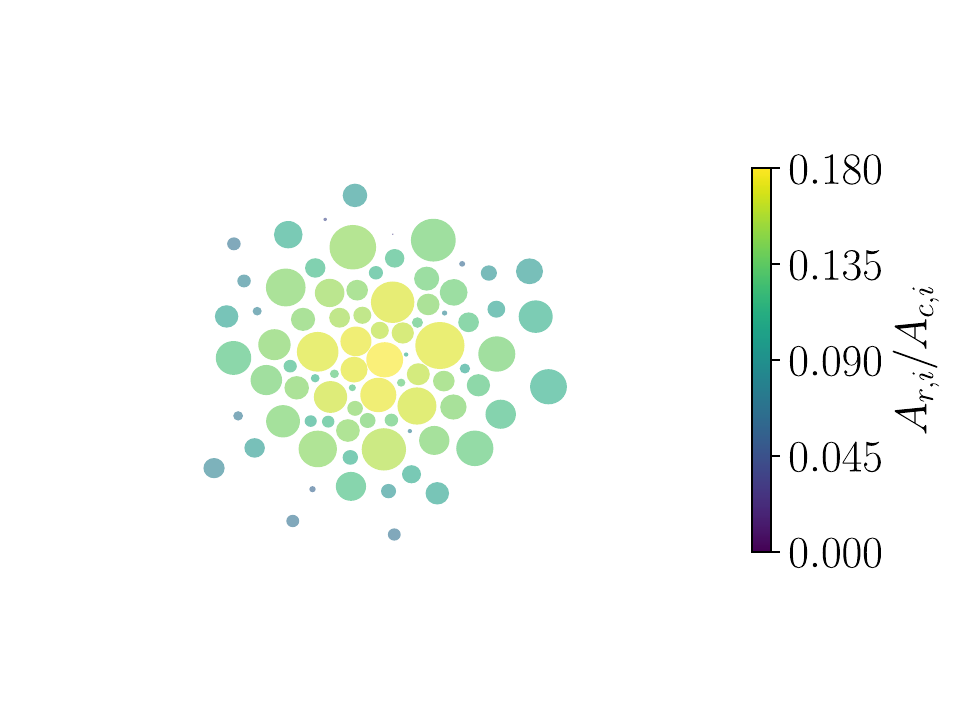}
}\\
\hspace*{-2cm} 
\subfloat[]{%
\includegraphics[scale=0.33]{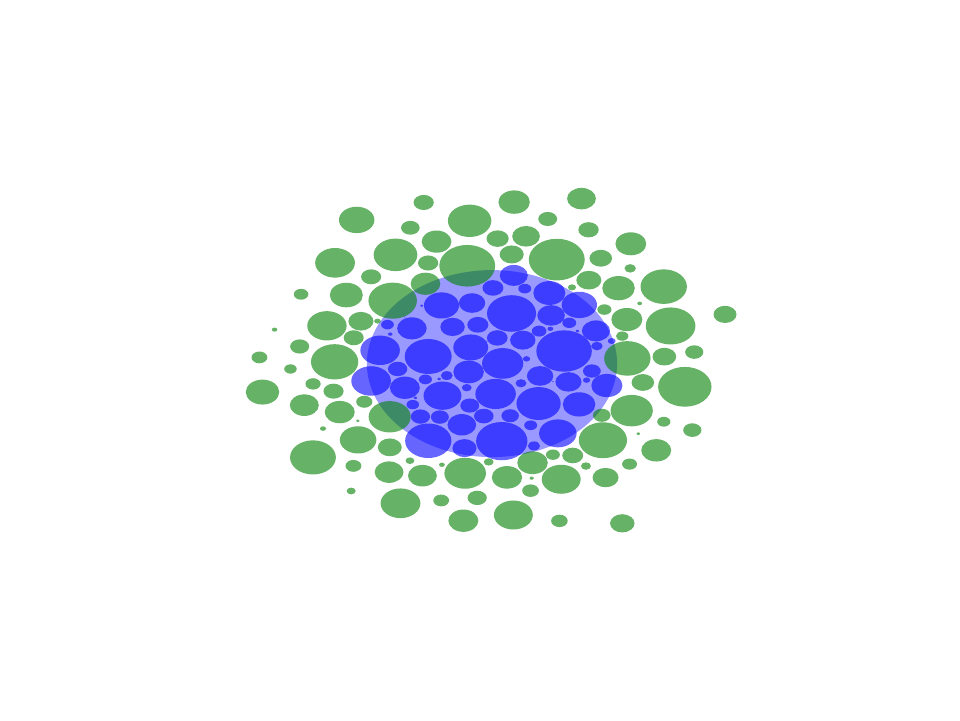}
\includegraphics[scale=0.33]{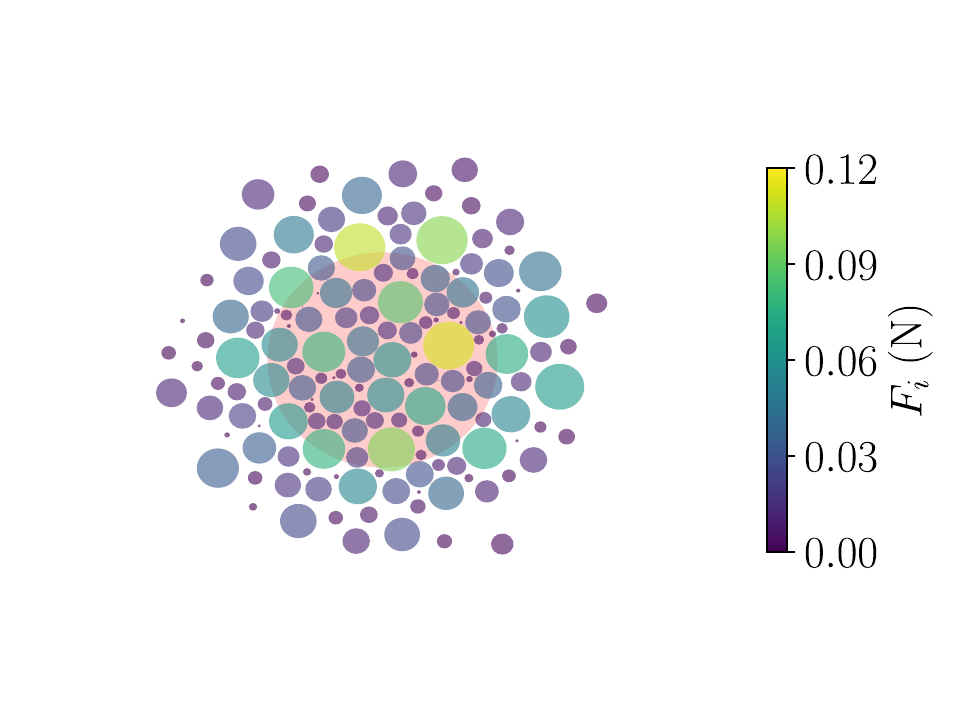}
\includegraphics[scale=0.33]{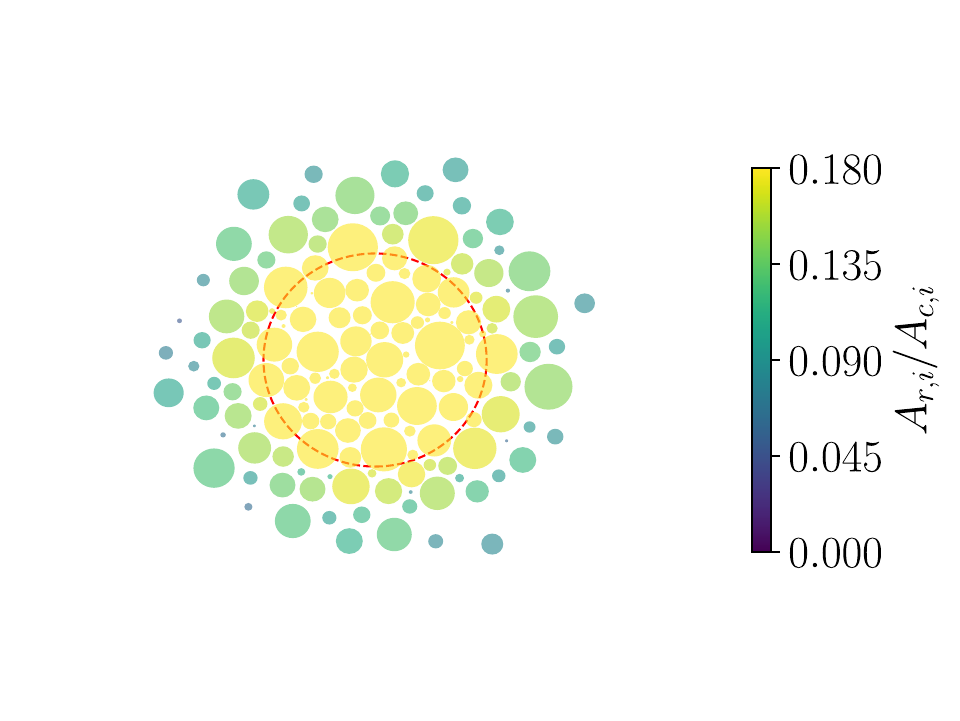}
}
\caption{
Contact mechanics model for (a) $d=0.7$ \textmu m and (b) $d=-2.4$ \textmu m, i.e., the contact area reaches the core particle (large circle).
The left snapshots give the contact patches, the middle snapshots the contact forces, and the right snapshots the real contact area of each macroscopic asperity according to \cref{eq:PersonAreaRatio}.
}
\label{fig:ForceMaps}
\end{figure*}

\Cref{fig:ForceMaps} visualizes the contact area for two different separation distances, (a) $d=0.7$ \textmu m and (b) $d=-2.4$ \textmu m.
A positive separation distance means that the contact plane is above the core particle. 
A negative separation distance means the contact area reaches the particle's core, and the asperities and the core particle form the contact area.
In the left snapshots, the green circles mark those macroscopic asperities that are in contact, and the blue circle marks the contact area of the core particle.
The middle snapshots depict the highly inhomogeneous contact force according to \cref{eq:Hertz_asp}.
In \Cref{fig:ForceMaps} (b), the red circle marks the area where the contact force of the core particle is added to the total contact force.

To sum up, \crefrange{eq:MactroContactArea}{eq:SumForce} provide the link between the separation distance $d$, the real contact area, $A_r$, and the total contact force, $F_n$.
Thus, these equations constitute a complete 'macroscopic' model for the real contact area of rough particles based on macroscopic asperities.

However, these macroscopic asperities represent a first-order approximation, whereas in real surfaces, the asperities are themselves also rough (cf.~\cref{fig:Topob}); 
in other words, the macroscopic asperities are covered by microscopic asperities.
Therefore, we test two approaches from the literature to further refine the model by accounting for the fractal surface.

The first approach by \citet{Pastewka2016} is based on simulations of randomly rough surfaces covered by spherical asperities.
They define the ratio of the real contact area of one macroscopic asperity to its contact area according to \cref{eq:MactroContactArea} as
\begin{equation} \label{eq:Pastewka}
\frac{A_{r,i}}{A_{c,i}}= \mathrm{erf} \left(\frac{\sqrt{\pi}\sigma_i}{2 p_\mathrm{rough}} \right) \, .
\end{equation}
Therein, $\sigma_i = F_{i}/A_{0}$, $p_\mathrm{rough} \equiv h^{'}_{\mathrm{rms}} E^*/\kappa$, $h^{'}_{\mathrm{rms}}$ is the root mean square slope of the surface, and $\kappa=2$ is a constant in the hard-wall limit.

The second tested approach by \citet{PERSSON2006201} defines the same ratio as
\begin{equation} \label{eq:PersonAreaRatio}
\frac{A_{r,i}}{A_{c,i}}= \mathrm{erf} \left(\frac{1}{\pi \sqrt{G_i}} \right)
\end{equation}
with
\begin{equation}
G_i=\frac{\pi \, E^*}{4 \, \sigma_i} \int^{\xi w_l}_{w_l}w^3 S(w) \ \mathrm{d}w \, .
\end{equation}
In the above equation, $S(w)$ is the surface roughness power spectrum according to \cref{eq:power}, $\xi$ is the maximum magnification, and $w_l$ is the largest wave vector which characterizes microscopic roughness.
The right column in \cref{fig:ForceMaps} plots \cref{eq:PersonAreaRatio}, thus, it shows the importance of microscopic roughness for the real contact area.

For both models of \citet{Pastewka2016} and \citet{PERSSON2006201}, the total real contact area is obtained analogously to \cref{eq:ar} by
\begin{equation}
A_r=\sum_{i=1}^N A_{r,i} \, .
\end{equation}

\Cref{fig:Impact_Area} compares the total contact area of a rough particle modeled by the approaches presented in this paper.
First, as a reference, the contact area of a perfectly smooth particle is analytically derived from the Hertz contact mechanics.
Second, the approach using only macroscopic asperities (\crefrange{eq:MactroContactArea}{eq:SumForce}) is illustrated.
Third, macroscopic asperities and additionally microscopic asperities by \cref{eq:Pastewka}, and fourth by \cref{eq:PersonAreaRatio} are presented.

To get the dependency of the contact area on the impact velocity, we solved Newton's second law of motion for the particle trajectory during the collision using a leapfrog integration scheme.

\subsection{Implementation of surface roughness in a charging model}
\label{sec:CT}

In the final step, we connect the contact mechanics and charge transfer models.
To this end, we chose the condenser model (cf.~\cref{eq:condenser}).
However, this choice is arbitrary; 
the contact mechanics model derived above is compatible with all charging models that depend on the contact area.

\begin{figure}[tb]
\centering
\subfloat[]{\includegraphics[width=0.5\textwidth]{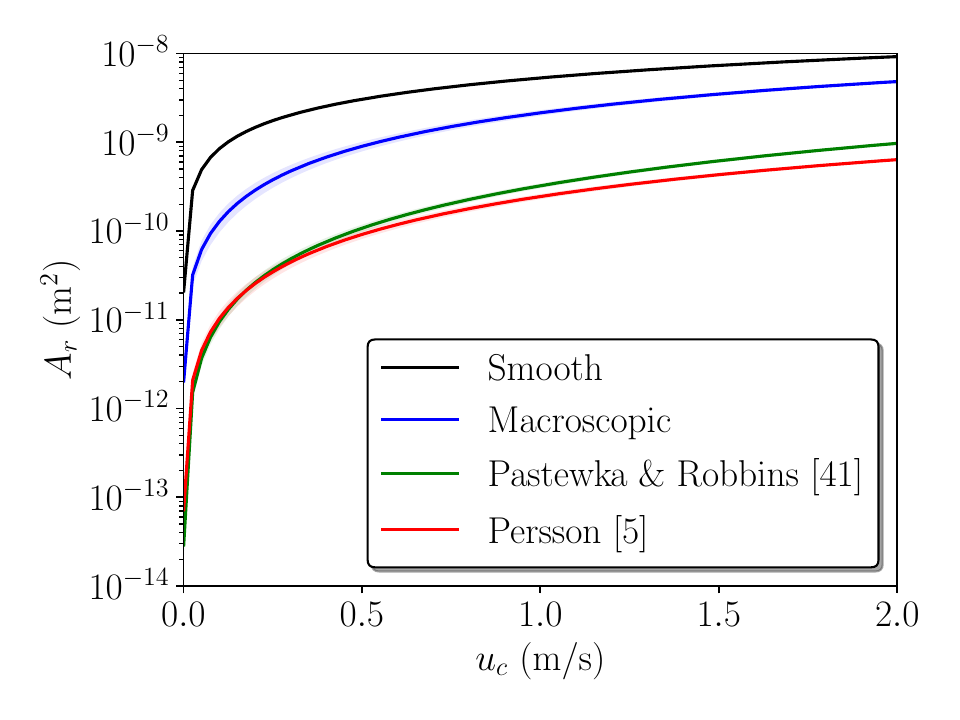}}
\subfloat[]{\includegraphics[width=0.5\textwidth]{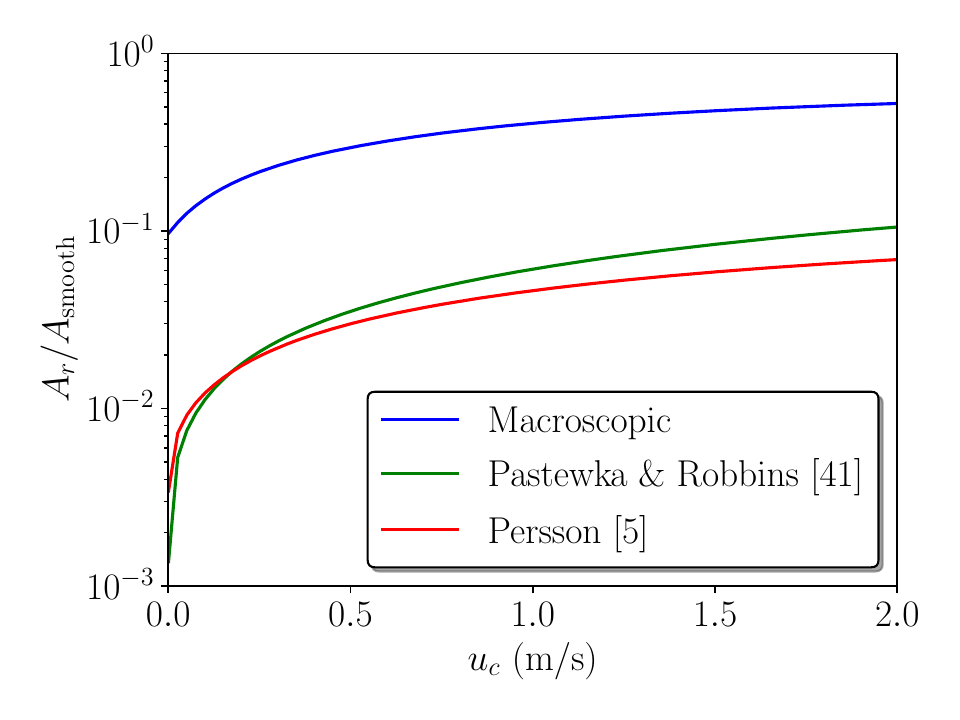}}
\caption{
(a) True total contact area during particle collision computed by four different approaches;
(i) assuming a smooth particle, (ii) modeling macroscopic asperities, and additionally modeling microscopic asperities using the approaches of (iii) \citet{Pastewka2016}, and of (iv) \citet{PERSSON2006201}.
(b) Ratio of the real contact area of the respective model to the Hertzian solution of the smooth sphere.
}
\label{fig:Impact_Area}
\end{figure}

The version of the condenser model employed assumes that the contact time is orders of magnitude shorter than the relaxation of charge across the particle, meaning the charge is transferred to the contact area and trapped there.
This assumption is valid for particles of high surface resistivity, i.e., insulator particles.
Therefore, we assume that the contact patch reaches its equilibrium charge on the timescale of the collision.
Then, \cref{eq:condenser} becomes
\begin{equation} \label{eq:localCondenser}
\Delta q = \beta V \varepsilon A_r/ z \, .
\end{equation}
As discussed in relation with \cref{eq:condenser}, for a given particle-target material system, all terms on the right-hand side are constant, except the real contact area.
Thus, $\Delta q \propto A_r$, which means that the curves in \cref{fig:Impact_Area} actually depict the impact charge divided by the constant factor $\beta V/z_c$.

\Cref{fig:Impact_Area} also highlights the error introduced by neglecting the influence of surface roughness. \Cref{fig:Impact_Area}(b) illustrates the ratio of the contact areas of a smooth to a rough sphere, representing the degree of underestimation. This error results in an overestimation of the contact area of 1/3 to 1/1000. In practical scenarios, however, one would use reference experiments of particles at impact velocities relevant to the process, inherently accounting for some surface roughness. Then, the contact area of a smooth sphere becomes a horizontal line in \cref{fig:Mc}, intersecting the model's prediction at the corresponding velocity. The error is then the difference between this line and the model's predictions. For example, if the experimental saturation charge is known at 0.91 m/s, the error for impact velocities of 0.5 m/s would be approximately 10\% and around 50\% at 0.25 m/s.

\section{Simulation of rough particle charging}
\label{sec:sim}

In this section, we present simulations of rough particle impact charging using the model derived in the above section, compare them to own experiments (\cref{sec:shaker}) and data from the literature (\cref{sec:plane}), and then we vary the surface roughness (\cref{sec:r}).

\subsection{Comparison with shaker experiments}
\label{sec:shaker}

Most published data on particle contact charging, such as the studies mentioned in the introduction \citep{Shuji_Matsusaka_2000,LESPRIT2021103605,WATANABE2007149}, do not state the particles' surface roughness or investigate soft particles and high impact velocities where the asperities deform completely.
Even though these studies confirm that the impact charge scales with the impact velocity, they do not inform on the effect of surface roughness on charging.

For comparison with the developed model, we conducted shaker experiments using particles for which the surface topology was analyzed in \cref{sec:Toposec}.
Details of the methodology of the shaker apparatus are described by \citet{JANTAC2019148}.
For each experiment, several thousand particles were put into a grounded stainless steel box and vertically shaken.
The ambient temperature during the experiments was 27~$^{\circ}$C, and the relative humidity was between 33\% and 43\%.
We recorded the shaking amplitude, frequency, and particle trajectories with a high-speed camera. 
The vertical shaking induces particle collisions with the box.
During these collisions, the particles charge until they asymptotically reach their saturation charge, $q_\infty$.
The experiment lasted long enough, so the measured particle charge after shaking corresponds to $q_\infty$.

\begin{figure}[tb]
\centering
\includegraphics[width=0.5\textwidth]{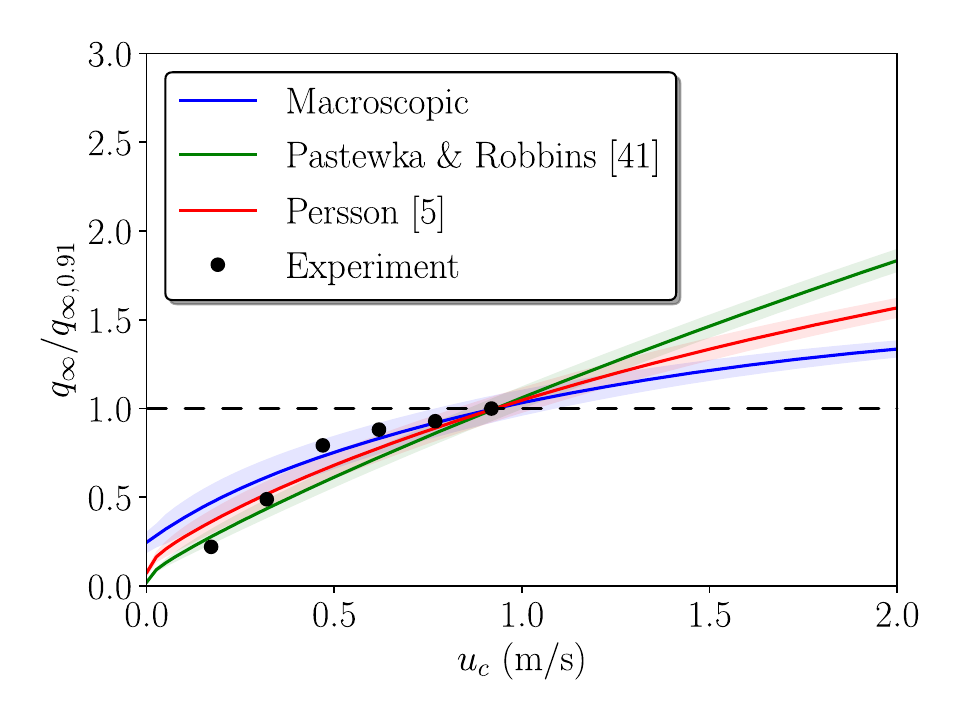} 
\caption{
Saturation charge simulation with three different models compared to results from shaker experiments.
The curves of the simulations give the mean of 2000 reconstructed surfaces and the shaded areas their standard deviations.
All saturation charges are normalized by the measured saturation charge at $u_c=0.91$~m/s. 
The dashed black line represents a prediction that ignores the effect of surface roughness, using a reference experiment conducted at $u_c=0.91$~m/s.
}
\label{fig:Mc}
\end{figure}

We varied the shaking amplitude and frequency between different experiments to obtain results for different impact velocities.
The mean and standard deviation of the impact velocity were extracted from the recorded particle trajectories. 
\Cref{fig:Mc} plots the experimentally measured saturation charge, normalized by the saturation charge at $u_c=0.91$~m/s, versus the mean of the impact velocity.

To simulate the saturation charge in dependence on impact velocity, we randomly reconstructed 2000 surfaces, each with the same surface roughness statistics as the measured particles.
The Young's modulus of the particles in the simulation was 500~MPa and their density 970~kg/m$^3$.
Each simulation provides the real contact area for one contact of the rough particle with the plane wall.
From this one contact, we extrapolate the real contact area of the complete particle surface $A_{r,tot}$, corresponding to the contact area that led to the measured saturation charge.
For the saturation charge \cref{eq:localCondenser} reads
\begin{equation}
\label{eq:totCondenser}
q_\infty = \beta V \varepsilon A_{r,tot}/ z_c \, .
\end{equation}
Using the experimentally measured saturation charge at $u_c=0.91$~m/s and the simulated real contact area of the total particle, we determine the constant parameters on the right-hand side of \cref{eq:totCondenser}.
Then, for each simulation, \cref{eq:totCondenser} returns the saturation charge for a given impact velocity.
\Cref{fig:Mc} plots the mean and the standard deviation of the saturation charge of the 2000 simulations.

In \cref{fig:Mc}, all three models agree well with the experiments in the considered data range.
The model predictions deviate more from each other for higher impact velocities, which were impossible to reach in the shaker.
Therefore, we cannot evaluate the model's accuracy beyond this limit. Ideally, we would have data points for normal collisions extending up to the range of plastic deformation, which represents the limit of our model for normal impact.

Small particles with low inertia slow down due to high aerodynamic drag.
Thus, also in real-world industrial facilities, particles rarely reach such high impact velocities.

Overall, we conclude from \cref{fig:Mc} that all models qualitatively predict the increase of the saturation charge with the impact velocity due to the related increase of the rough contact area.
Further, all models predict the saturation charge with sufficient accuracy.
However, the macroscopic model advantage over other models is that it can reconstruct the contact surface with a limited number of statistical parameters.
In particular, the statistical model reconstructs surfaces without the root mean square slope of the surface ($h^{'}_{\mathrm{rms}}$) in \cref{eq:Pastewka} or the surface roughness power spectrum ($S(w)$ in \cref{eq:PersonAreaRatio}).
The input data to the macroscopic model are standard metrics of surface roughness.

Currently, the model takes into account the impact velocity. 
Simulating different environmental conditions requires experimental data at those conditions to determine the factor $\beta V \varepsilon / z_c$.


\subsection{Plane surface charging}
\label{sec:plane}

\begin{figure}[tb]
\centering
\includegraphics[width=0.5\textwidth]{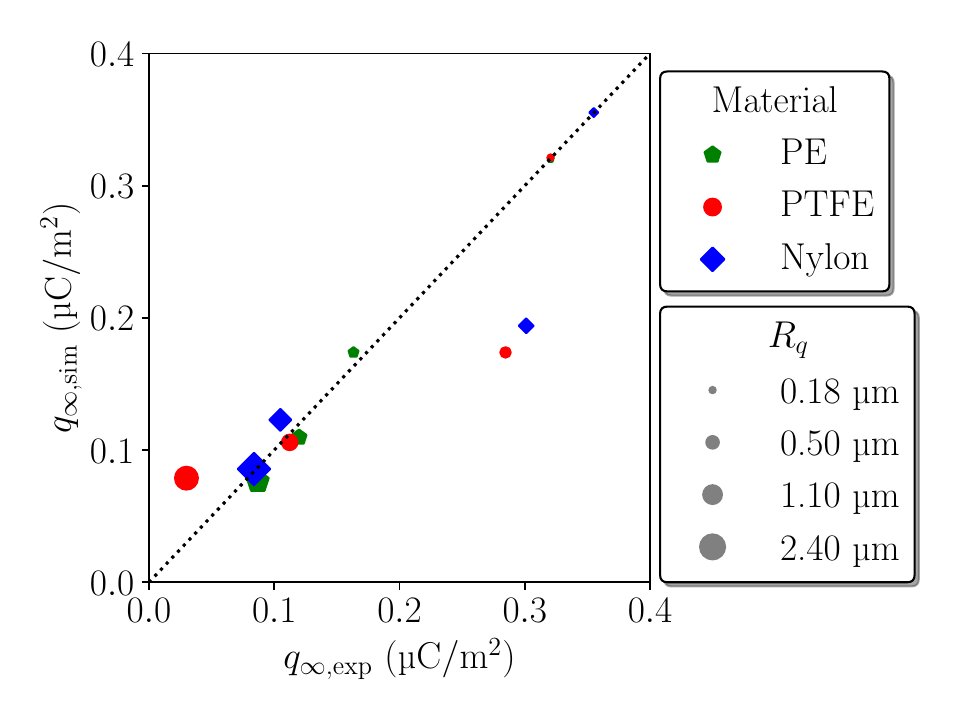}
\caption{
Parity plot of saturation charges for different materials and varying surface roughness.
The data point's size indicates the surface roughness.
Experimental data were obtained from \citep{WANG2019};
the simulations used the macroscopic model.
The constants of the model were determined from the measured saturation charge of the smoothest surface ($R_\mathrm{q}= 0.18 \ $\textmu m), where the saturation charge densities of PE and PTFE coincide.
}
\label{fig:RouggnessParity}
\end{figure}

In the absence of literature data on the charging of particles with defined surface roughness, we compare our model to the data of \citet{WANG2019} of the charging of plane surfaces.
They pressed a plane glass surface in the normal direction against other plane surfaces of four different materials: PE, PTFE (polytetrafluoroethylene), nylon, and aluminum.
Further, they varied and reported these materials' surface roughness ($R_q$).
After separation, they measured the charge on the glass.
We omitted data where the charge polarity switched from the comparison because the condenser model can only predict net charge transfer in one direction.

Since higher-level data of the surface topology (e.g., $h^{'}_{\mathrm{rms}}$ or $S(w)$) are not reported, we used the macroscopic model to simulate this experiment.
To adjust to the experimental setting, we seeded the asperities not on a curved but a flat surface.
Since the contact surface is flat, the measured and simulated charges directly translate to surface charge densities with the units of C/$\mathrm{m}^{2}$.

We assumed that the contact load of each experiment was constant and the contact force was applied in the normal direction, although the authors stated that some sliding motion of the surfaces was unavoidable.
Further, they contacted and held for approximately five seconds, from which we infer that the contact was in mechanical equilibrium and that the contact load was the sum of the weights of the test surface and the device to which it was attached.

As in the above section, we determined the constants on the right-hand side of \cref{eq:totCondenser} from one experimental condition.
Here, we chose for each material pair the experimental saturation charge of the smoothest reported surface, i.e., $R_q=0.18$ \textmu m.

\Cref{fig:RouggnessParity} shows the parity plot between the experimental~\citep{WANG2019} and simulated saturation charges.
The simulated data is the average over 500 rough surfaces for each $R_q$.
The experiments and simulations coincide for the smoothest surface since these experimental data were used as a reference for the model constants.
Overall, the experiments and simulations agree well, especially for PE.
For PTFE and Nylon, the model underestimates the saturation charge for $R_q=0.5$ \textmu m but agrees well with experiments on rougher surfaces.
    We note that \citet{WANG2019} does not mention an estimated error for this type of experiment. Therefore, we cannot estimate if the error at $R_q=0.5$ \textmu m is caused by our approach or by the experiment.

In summary, this section shows that our approach with macroscopic asperities can model the effect of surface roughness on the saturation charge.

\subsection{Varying surface roughness}
\label{sec:r}

\begin{figure}[tb]
\centering
\subfloat[]{%
\label{fig:r-u}%
\includegraphics[width=0.47\textwidth]{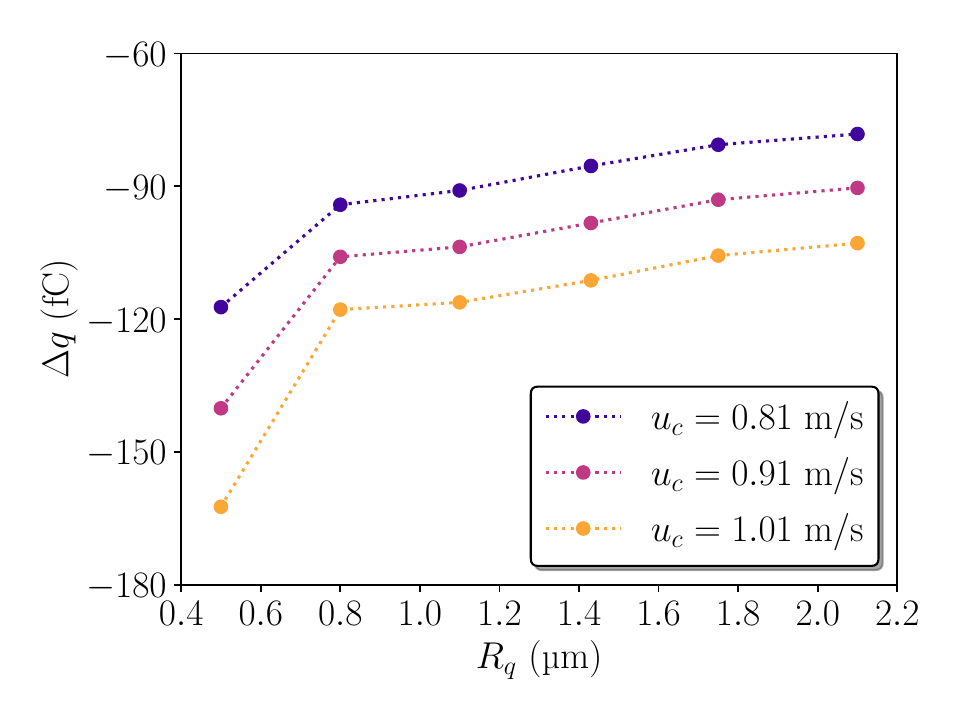}
}
\subfloat[]{
\label{fig:r-d}
\includegraphics[width=0.47\textwidth]{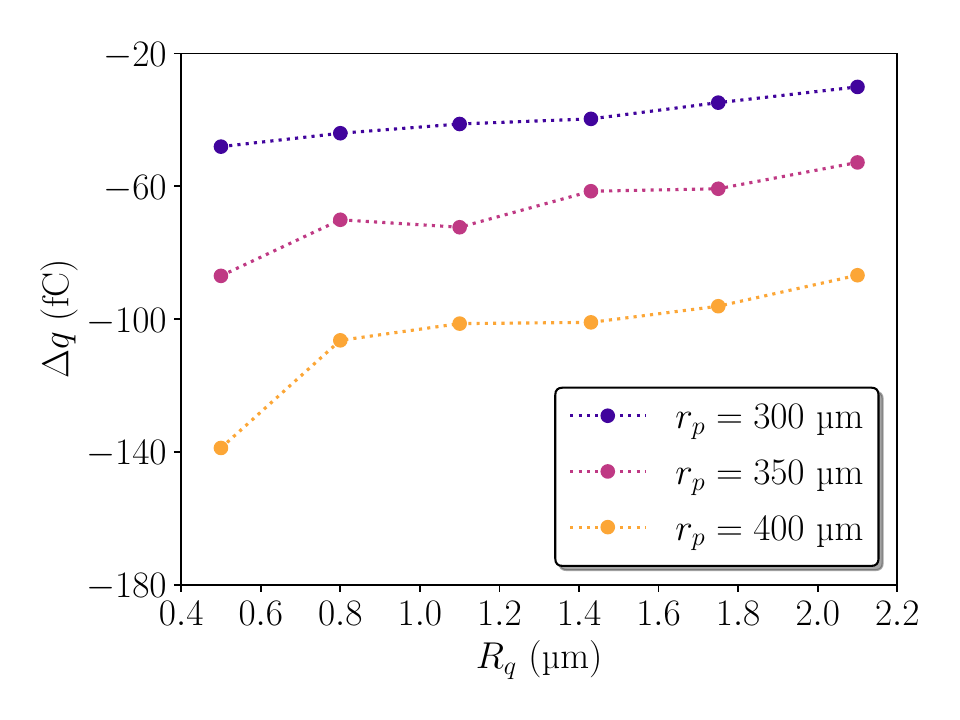}
}
\caption{
Impact charge ($\Delta q$) of uncharged particles of different surface roughness.
(a) For $r_p= 400$ \textmu m and varying impact velocity.
(b) For impact velocity $u_c=0.91$ m/s and varying particle size.
}
\label{fig:r}
\end{figure}

In the final section, we simulate the effect of surface roughness on the impact charge.
To this end, we model the particles of \cref{sec:shaker} and use the obtained constants ($\beta V \varepsilon / z_c$) in \cref{eq:localCondenser}.

All results in \cref{fig:r} demonstrate that increasing the surface roughness decreases the magnitude of the impact charge (note that the impact charges are negative).
Deforming rougher surfaces requires more impact energy than smoother surfaces.
Therefore, the real contact area of rough surfaces is smaller than that of smooth surfaces.
Consequently, smooth particles obtain a higher impact charge for the same impact conditions than rough particles.

According to \cref{fig:r-u}, increasing the impact velocity increases the impact charge.
The simulated particles have a radius of 400 \textmu m and are uncharged before the impact.
Increasing the impact velocity yields higher kinetic energy.
Dissipating the kinetic energy during impact results in stronger asperity deformation and an increase of the real contact area and the impact charge.

Further, according to \cref{fig:r-d}, increasing the particle radius increases the impact charge.
Particles of larger size have a higher kinetic energy, which, similarly to a higher impact velocity, results in a larger real contact area and higher impact charge.

\section{Conclusions}
\label{sec:con}
In this paper, we developed a model for triboelectric particle charging that accounts for the particle surface roughness.
Under the assumption that charge transfer scales with the real contact area, which holds especially for insulating surfaces, the model predicts the dependence of the impact charge on surface roughness and the elastic deformation of surface asperities under pressure.
The model requires only standard metrics of the surface roughness ($R_a$, $R_q$) as input information regarding the particles' topology.
Due to the simplicity and availability of these standard metrics, the model appears to be suited for potential implementation in large-scale simulations of the electrification of particle-laden flows.


%

\section*{Author contributions}
\textbf{Simon Jantač}: Conceptualization, Methodology, Software, Validation, Formal analysis, Investigation, Data Curation, Writing - Original Draft, Writing - Review \& Editing, Visualization. \textbf{Jarmila Pelcová}:  Validation.
\textbf{Jana Sklenářová}: Validation, Investigation.
\textbf{Marek Drapela }: Validation.
\textbf{Holger Grosshans}: Writing - Review \& Editing,  Funding acquisition, Supervision.
\textbf{Juraj Kosek}: Writing - Review \& Editing,  Funding acquisition, Supervision

\section*{Acknowledgement}

SJ and HG received funding from the European Research Council~(ERC) under the European Union's Horizon 2020 research and innovation program~(grant agreement No.~947606 PowFEct).
The work of JP and JS forms part of the research program of DPI, project \#830.

\section*{Apendix}
\appendix
\section{Spheroid asperties}
\label{sec_A1}
A more general approach is to use spheroids instead of spheres to approximate the shape of asperities.
We can substitute spheres with spheroids by adding one semi-axis, i.e., we define each asperity by the position of its center, $x_i$, $y_i$, and $z_i$ in Cartesian or $\theta_i$ and $\phi_i$ in spherical coordinates.
The asperity's size is given by its semi-axes $a_i$, and $b_i$.
That means the spheroid is oblate if $b_i > a_i$ or prolate if $b_i < a_i$.
In the statistical reconstruction of the surface, we treat $b_i$ of the spheroid as $r_i$ in the case of a sphere.
A linear model could be used instead since the Hertz model is not analytically derived for spheroids.
Here, the contact area at penetration depth $\delta_i$ is defined according to~\citep{Calvimontes2018}
\begin{equation}
A_i=\pi b_i^2 \left[ 1-\left(1-\frac{\delta_i}{a_i} \right)^2 \right]. 
\end{equation}
The same equation calculates the contact area in the triboelectric model.

\bibliographystyle{elsarticle-num-names} 
\bibliography{References}

\end{document}